\documentclass[prl,aps,showpacs,preprintnumbers,amsmath,amssymb,nofootinbib,showkeys,twocolumn,floatfix]{revtex4-1}

\pdfoutput=1
\usepackage{xcolor}
\usepackage{hyperref}
\usepackage{natbib}
\usepackage{subfiles}
\usepackage{aas_macros}
\usepackage{amssymb}
\usepackage{graphicx}% Include figure files
\usepackage{dcolumn}% Align table columns on decimal point
\usepackage{bm}% bold math
\usepackage{cases}
\usepackage[lofdepth,lotdepth,caption=false]{subfig}

\newcommand{\vesc}{v_{esc}}
\newcommand{\Rstar}{R_{\star}}
\newcommand{\Mstar}{M_{\star}}
\newcommand{\be}{\begin{equation}}
\newcommand{\ee}{\end{equation}}
\newcommand{\pn}{p_N(\tau)}

\newcommand{\avg}[1]{\langle #1 \rangle}

\newcommand{\unit}[1]{\mathrm{#1}}
\newcommand{\GeV}{\unit{GeV}}

\newcommand{\percc}{\unit{cm}^{-3}}
\newcommand{\Msun}{M_{\odot}}

\begin{document}
\title{Probing below the neutrino floor with the first generation of stars}

\author{Cosmin Ilie}
\email[E-mail at: ]{cilie@colgate.edu}
 \altaffiliation[Additional Affiliation: ]{Department of Theoretical Physics, National Institute for Physics and Nuclear Engineering,  Magurele, P.O.Box M.G. 6, Romania}%Lines break automatically or can be forced with \\
\author{Caleb Levy}
\affiliation{ Department of Physics and Astronomy, Colgate University\\
13 Oak Dr., Hamilton, NY 13346, U.S.A.
%\\This line break forced with \textbackslash\textbackslash
}%
\author{Jacob Pilawa}%
\affiliation{Department of Astronomy, University of California at Berkeley\\
Berkeley, CA 94720 U.S.A.}
\author{Saiyang Zhang}
\affiliation{Theory Group, Department of Physics, University of Texas\\
Austin, TX 78712, U.S.A.}

\date{\today}%

\begin{abstract}
    We show that the mere observation of the first stars (Pop~III stars)  in the universe can be used to place tight constraints on the strength of the interaction between dark matter and regular, baryonic matter. We apply this technique to a candidate Pop~III stellar complex discovered with the Hubble Space Telescope at $z\sim7$ and find some of the deepest bounds to-date for both spin-dependent and spin-independent DM-nucleon interactions, over a large swath of DM particle masses. Additionally, we show that the most massive Pop~III stars could be used to bypass the main limitations of direct detection experiments: the neutrino background to which they will be soon sensitive. 
\end{abstract}

\maketitle
One of the most intriguing open questions in Physics today is the nature of Dark Matter (DM). Its existence has been inferred via the gravitational effects it has from the smallest scales, in the Cosmic Microwave Background (CMB) radiation~\cite{Komatsu:2009,Komatsu:2011,Ade:2015,Aghanim:2018}, to intermediate, galactic~\cite{Rubin:1970} and cluster scales~\cite{zwicky1937masses}. The complex large scale structures and sub-structures dark matter forms during its gravitational collapse around the potential wells provided by the primordial density fluctuations can be mapped using gravitational lensing on galactic cluster~\cite{Natarajan:2017} and cosmological~\cite{Madhavacheril:2014,Vikram:2015,Hikage:2018} scales. Over the past few decades a standard, concordance cosmological model has emerged as a leading candidate that best explains all available cosmological data: the $\Lambda$-CDM model. About $27\%$ of the energy budget of the Universe today is in the form of DM, whereas regular, baryonic matter only amounts to roughly $5\%$. The other $68\%$ is thought to be comprised of Dark Energy: a uniform, negative pressure fluid responsible for the current accelerated expansion of the Universe.  

{\it{Dark Matter detection.   }}Currently, there are three broad strategies in the hunt for dark matter: production of DM particles in accelerators, direct, and indirect detection. Each of these exploits the various possible interaction channels between dark matter and baryonic matter. The Large Hadron Collider (LHC) has not yet found any evidence of particles outside of the standard model; as such, the minimum mass of any supersymmetric DM particle candidate has been pushed to higher and higher values. Indirect detection experiments rely on the possible self-annihilations or decay of dark matter particles, whenever DM densities are high. The nearest such site is the center of our own galaxy. An antiproton and a gamma-ray excess compared to known backgrounds have  been found in Alpha Magnetic Spectrometer (AMS) and Fermi satellite data, respectively. Intriguingly, both of those excesses could be fit with a $\sim60~\unit{GeV}$ DM particle self-annihilating~\cite{Goodenough:2009,hooper2011dark,Cholis:2019}. Alternatively, there are astrophysical explanations for those excesses~\cite{Gordon:2013,YUAN:2014,Kohri:2015}.

Direct detection experiments exploit the small amount of energy a DM particle deposits as it collides with atomic nuclei~\cite{Goodman:1985,Drukier:1986}. As such, they are extremely challenging; moreover, shielding from overwhelming cosmic ray backgrounds requires performing the experiments deep underground. So far, DAMA/LIBRA is the only group that reports a signal consistent with DM detection~\cite{BERNABEI:1998,Bernabei:2014,Bernabei:2018}. In lack of a clear detection signal, direct detection experiments are constraining the allowed strength of the interaction between dark matter and baryonic matter. As they become more and more sensitive, their detectors will be swamped with signals from neutrinos, which cannot be disambiguated from any possible dark matter signal. At that stage, if no clear DM signal identification has been made, new detection strategies will need to be implemented. For reviews on dark matter and its detection status see Refs.~\cite{Bertone:2005,Bertone:2016,Freese:2017dm,Gaskins:2016,Lin:2019,losHeros:2020}.

In this  letter, we propose a novel method of constraining the dark matter proton scattering cross section using Pop~III stars, applicable when DM self-annihilates. Using this formalism for the candidate Pop~III system at redshift $z\sim7$, found in the Hubble Space Telescope (HST) data~\cite{Vanzella:2020}, we obtain the most stringent bounds to-date for DM with masses outside of the commonly explored WIMP window. We point out that the upcoming James Webb Space Telescope (JWST), and its potential for discovering massive Pop~III stars, could be used to probe below the neutrino floor that will soon limit direct detection experiments on earth. 

{\it{Method.   }}Pop~III stars formed via the gravitational collapse of zero metallicity, primordial baryonic
gas clouds that contain pristine H and He from big bang nucleosynthesis. This happened at high redshifts ($z\sim10-50$)~\footnote{Sometimes at $z$ as low as $~7$~\cite{Mebane:2018}}, at the center of DM mini-halos ($M_{halo}\sim 10^6\Msun$), in very DM-rich environments. Using hydrodynamical simulations, the following picture emerges: typically one or just a few Pop~III stars form per mini-halo, within the inner 10 AU of the center, with masses up to $\sim 1000\Msun$, powered by $H$ fusion~\cite{Barkana:2000,Abel:2001,Bromm:2003,Yoshida:2006,Yoshida:2008,Loeb:2010,Grief:2012,Bromm:2013,Klessen:2018}. Under certain conditions~\cite{Spolyar:2008dark}, DM heating during the formation of the first stars leads to objects powered by DM annihilations, Dark Stars (DS). Those hypothetical objects can grow to be supermassive~\cite{Freese:2010smds} and have different photometric signatures compared to Pop~III stars~\cite{Zackrisson:2011,Ilie:2012}. In this work we assume that Dark Stars and Pop~III stars are not mutually exclusive, and that at least some of the first stars are Pop~III stars, which are the probes we use to constrain DM.

Any astrophysical object can accrete dark matter at its core via a phenomenon called capture~\cite{Press:1985,Gould:1987,Gould:1987resonant}. Pop~III stars, forming in a DM-rich environment, are particularly good probes of this phenomenon. Refs.~\cite{Freese:2008cap,Iocco:2008cap} study this for weakly interacting (WIMP) dark matter that gets captured by at most one collision (single-scattering) with nuclei inside Pop~III stars. Using the recently developed multiscattering capture formalism~\cite{Bramante:2017}~\footnote{See also~\cite{Ilie:2020Comment}.}, two of the authors of this letter investigated the capture of superheavy ($m_X\gtrsim 10^8~\unit{GeV}$) dark matter by Pop~III stars~\cite{Ilie:2019}, finding that heating from dark matter annihilations leads to an upper bound on the Pop~III masses. In this letter, we show how the mere observation of a Pop~III star of any given mass can be used to constrain the DM-proton scattering cross section. Any star of a given mass can never shine brighter than the Eddington luminosity ($L_{Edd}$):
\be\label{eq:Eddington}
L_{nuc}(\Mstar)+L_{cap}(\Mstar; {{\text{DM params}}})\leq L_{Edd}(\Mstar).
\ee
$L_{nuc}$ represents the heating generated by the hydrogen fusion at the core of the star, whereas $L_{cap}$ is the heating due to captured dark matter annihilations, which depends both on stellar~\footnote{Homology relations relate $\Rstar$ with $\Mstar$~\cite{Ilie:2020PopIIIa}} and DM parameters. Most importantly, it is sensitive to the DM-proton scattering cross section ($\sigma$). This ultimately allows us to place bounds on $\sigma$, if all other parameters are measured or constrained. Conversely, if $\sigma$ is known, or constrained by other experiments, we can place upper bounds on the mass of Pop~III stars, since the Eddington limit scales linearly with mass, and $L_{nuc}\sim\Mstar^p$ and $L_{cap}\sim\Mstar^m$, with both $p$ and $m$ greater than one. 

Below we continue the discussion of our method, explaining how we calculate  $L_{cap}, L_{nuc}$, and $L_{Edd}$. DM particles crossing a star with radius $\Rstar$ can, via collisions with nuclei, lose enough energy to become trapped by the gravitational field of the star. This happens when the DM particle velocity falls below the escape velocity ($\vesc$) of the star. The capture rate is given by ~\cite{Bramante:2017}:
\begin{widetext}
\be
\label{Eq:Ctot}
C_{tot} = \sum_{N=1}^{\infty} C_{N} = \sum_{N=1}^{\infty} \underbrace{\pi \Rstar^2}_\textrm{capture area}\times \,\underbrace{\frac{\rho_X}{m_X} \int_0^{u_{max;N}} \dfrac{f(u)du}{u}\,(u^2+v_{ esc}^2)}_\textrm{flux of DM captured after N collisions}\times \, \underbrace{p_{ N}(\tau)}_\textrm{prob. for $N$ collisions}.
\ee
\end{widetext}
The probability a DM particle will collide exactly $N$ times as it crosses the star has the following closed form~\cite{Ilie:2019}: $\pn=\frac{2}{\tau^2}\left(N+1-\frac{\Gamma(N+2,\tau)}{N!}\right)$, where $\Gamma(a,b)$ is the incomplete gamma function. The optical depth is defined as: $\tau=2\Rstar~\sigma~n_T$, where $n_T$ is the average number density of nuclei inside the star. Throughout, $\rho_X$ represents the DM density.  DM particles with velocity $u$ (measured infinitely far from the star) greater than $u_{max;N}=\vesc\left[(1-\beta_+/2)^{-N}-1\right]^{1/2}$ will not be captured after $N$ collisions, since they are too fast to be slowed below $\vesc$. Here, $\beta_+\equiv 4mm_X/(m+m_X)^2$, with $m$ being the mass of the target nucleus. For this reason, we only integrate the velocity distribution up to the $u_{max}$ cutoff. This amounts to only a part of the DM particle flux crossing the star being captured. The key point is that, as expected, the capture rate depends on the scattering cross section (via $\tau$). In~\cite{Ilie:2020Comment,Ilie:2020PopIIIa}, we found closed form analytical expressions for $C_{tot}$, obtained  assuming a Maxwell-Boltzmann distribution, $f(u)$.

Via collisions with nuclei, captured DM particles will thermalize, following a truncated Maxwellian distribution, since there are no captured particles with $w>v_{esc}$. In the limit of weak cross section ($\sigma$), which is applicable here, the DM temperature, $T_{X}$, becomes constant throughout the star and can be directly related to the central baryon temperature of the star $T_c$~\cite{Spergel:1985}. In~\cite{Ilie:2020PopIIIa}, we find that this temperature ranges from $0.6 T_c$ (if $m_X\lesssim 10^{-2}~\unit{GeV}$) to $T_c$ (if $m_X\gtrsim 1~\unit{GeV}$). DM will follow an isothermal sphere profile, i.e. $n_X=n_{X,c} e^{-m_X\Phi/T_X}$, with $\Phi$ being the gravitational potential inside the star. 

For sub-GeV DM particles, ``evaporation'' (i.e. the rate at which captured DM particles escape the star from up-scattering to velocities above the escape velocity via collisions with nuclei) becomes important. Pop~III stars are radiation pressure dominated and, as such, can be approximated by an $n=3$ polytrope. In ~\cite{Ilie:2020PopIIIa} we find:
\be\label{eq:EvapRate}
E\approx\frac{3V_{\star}\bar{n}_pu_c\sigma}{2V_1\sqrt{\pi}}e^{-\frac{v_{esc}^2\mu}{u_c^2\Theta}(1+\xi_1/2)},
\ee
where $V_{\star}$ represents the volume of the star, $\bar{n}_p$ is the average proton number density, $u_c\equiv\sqrt{\frac{2T_c}{m_p}}$, i.e. the average thermal velocity of protons at the center of the star, $\mu\equiv m_X/m_p$, $\Theta\equiv T_X/T_c$, and $\xi_1\approx6.89$ is the first node of the Lane-Emden function for $n=3$. Additionally, $V_j\equiv\int_{\star}dVe^{-jm_X\Phi/T_X}$.

 Assuming DM self-annihilates, the number of DM particles inside the star is controlled by the following:
 \be
 \dot{N}_{X}=C_{tot}-C_A N_X^j-EN_X,
 \ee
 with $C_A$ being an $N_X$ independent annihilation coefficient, and $j$ the number of DM particles entering each annihilation process. In this letter, we consider four different annihilation mechanisms: the standard p/s-wave annihilations ($j=2$), relevant for DM particles with masses above a few GeVs, and Strongly Interacting Massive Particles (SIMP) ($j=3$)~\cite{Hochberg:2014} and Co-SIMP ($j=2$)~\cite{Smirnov:2020} models, which have freeze-out annihilation cross sections $\avg{\sigma v^2}$ that naturally explain production of sub-GeV DM to match the observed relic abundance. The annihilation coefficient is given by: $C_A^{2\to2}=\int_{\star}dV n_X^2\avg{\sigma v}/(\int_{\star}dV n_X)^2$  (p/s-wave), $C_A^{Co-SIMP}=\int_{\star}dV n_X^2n_{SM}\avg{\sigma v^2}/(\int_{\star}dV n_X)^2$ (Co-SIMP DM, where $n_{SM}$ is the baryon number density), and $C_A^{SIMP}=\int_{\star}dV n_X^3\avg{\sigma v^2}/(\int_{\star}dV n_X)^3$  (SIMP DM). In the absence of evaporation, the capture and annihilation processes equilibrate in a timescale $\tau_{eq}\equiv(C_{tot}C_A)^{-1/2}$.
 
Evaporation shortens the equilibration timescale: $\tau_{eq}\to\tau_{eq}/\kappa$ (with $\kappa\equiv\sqrt{1+E^2\tau_{eq}^2/4}$). At larger times, $N_X$ attains a constant value. We find that the SIMP DM process (DM+DM+DM$\to$DM+DM) is inefficient in equilibrating the capture and annihilation/evaporation processes~\cite{Ilie:2020PopIIIa}. Conversely, all the other three processes of interest: p/s-wave (DM+DM$\to$SM+SM) and Co-SIMP (DM+DM+SM$\to$DM+SM) lead to equilibration time scales much smaller than the lifetime of the star~\cite{Ilie:2020PopIII}. 
 
 As DM particles enter an equilibrium regime, the annihilation rate $\Gamma_A=C_AN_X^2$ ($j=2$ s/p-wave and Co-SIMP DM) becomes constant, and can be related to the capture and evaporation rates:  $\Gamma_A=C_{tot}/(\kappa+1/2E\tau_{eq})^2$. Assuming a fraction $f$~\footnote{$f$ is a model-dependent, order unity number. We assume $f=1$.} of the energy from the annihilation products gets thermalized inside the star, we  can express the luminosity due to captured DM heating as: $L_{cap}=fC_{tot}m_X/(\kappa+1/2E\tau_{eq})^2$, using the $C_{tot}$ numerically calculated from Eq.~\ref{Eq:Ctot}. Alternatively, in~\cite{Ilie:2020Comment,Ilie:2020PopIIIa} we introduced analytic approximations for $C_{tot}$ that can cover the entire $(m_X,\sigma)$ parameter space.

In order to exploit the sub-Eddington condition (Eq.~\ref{eq:Eddington}), we also need the nuclear luminosity ($L_{nuc}$) and the Eddington limit ($L_{Edd}$). For $L_{nuc}$, we find the following interpolating function:
\begin{equation}\label{eq:Lnuc}
L_{n u c} \simeq 10^{\frac{\log \left(3.71 \times 10^{4} L_{\odot} \mathrm{s} / \mathrm{erg}\right)}{1+\exp (-0.85 x-1.95)}} \cdot x^{\frac{2.00}{x^{0.48}+1}} \operatorname{erg} / \mathrm{s}
\end{equation}
where $x \equiv \frac{M_\star}{M_\odot}$ and $L_\odot \equiv 3.846 \times 10^{33} \operatorname{erg} / \mathrm{s}$.
As expected, this logistic fit function transitions between $L_{nuc}\sim\Mstar^3$ for intermediate mass stars, to $L_{nuc}\sim\Mstar$, for $\Mstar\gtrsim 1000\Msun$. For the Eddington luminosity, assuming BBN composition of Pop~III stars, we find: $L_{Edd} \approx 3.71\times 10^{4} (M_\star/M_\odot)L_\odot.$

 We sum up our method: by using the sub-Eddington condition (Eq.~(\ref{eq:Eddington})) we can find an upper bound on $\Mstar$ for Pop~III stars, when $\sigma$ is constrained via direct detection experiments. Conversely, once a Pop~III star with a given mass is identified, we can use that information to place constraints on $\sigma$ as a function of $m_X$. In this letter, we will use the Pop~III stellar system candidate found by~\cite{Vanzella:2020} to place some of the most stringent constraints on $\sigma$.
 
 {\it{Dark Matter Densities.   }} Our bounds are sensitive to the ambient DM density at the location of the Pop~III star. Since $\rho_X$ is not directly constrained, this is the main limitation of our method. In this section, we discuss the range of ambient DM densities considered throughout our work, and explain the rationale behind our choice. We point out that all of our bounds are inversely proportional to $\rho_X$. Therefore, it is straightforward to re-scale our bounds to any other assumed value of $\rho_X$.
 
 We assume an adiabatically contracted Navarro-Frenk-White (NFW) DM profile for the host minihalo. As the baryonic protostellar cloud cools and collapses, it will modify the initial DM density profile by enhancing densities in the inner regions of the halo. This is simply a response of the DM orbits to an increase in the gravitational potential. The Adiabatic Contraction (AC) formalism~\cite{Blumenthal:1985,Young:1980,Gendin:2004,Gendin:2011} can be used to estimate this DM density enhancement, using the simplifying assumption of the existence of adiabatic invariants for DM particles inside a halo. Results from numerical simulations are in good agreement with those obtained via the adiabatic contraction formalism~\cite{Sellwood:2005,Gendin:2011}, especially for high redshift halos, such as those where Pop~III stars form, since baryonic feedback effects are not important in this case. 
 \begin{figure} [bht]
    \centering
    \includegraphics[width=0.99\linewidth]{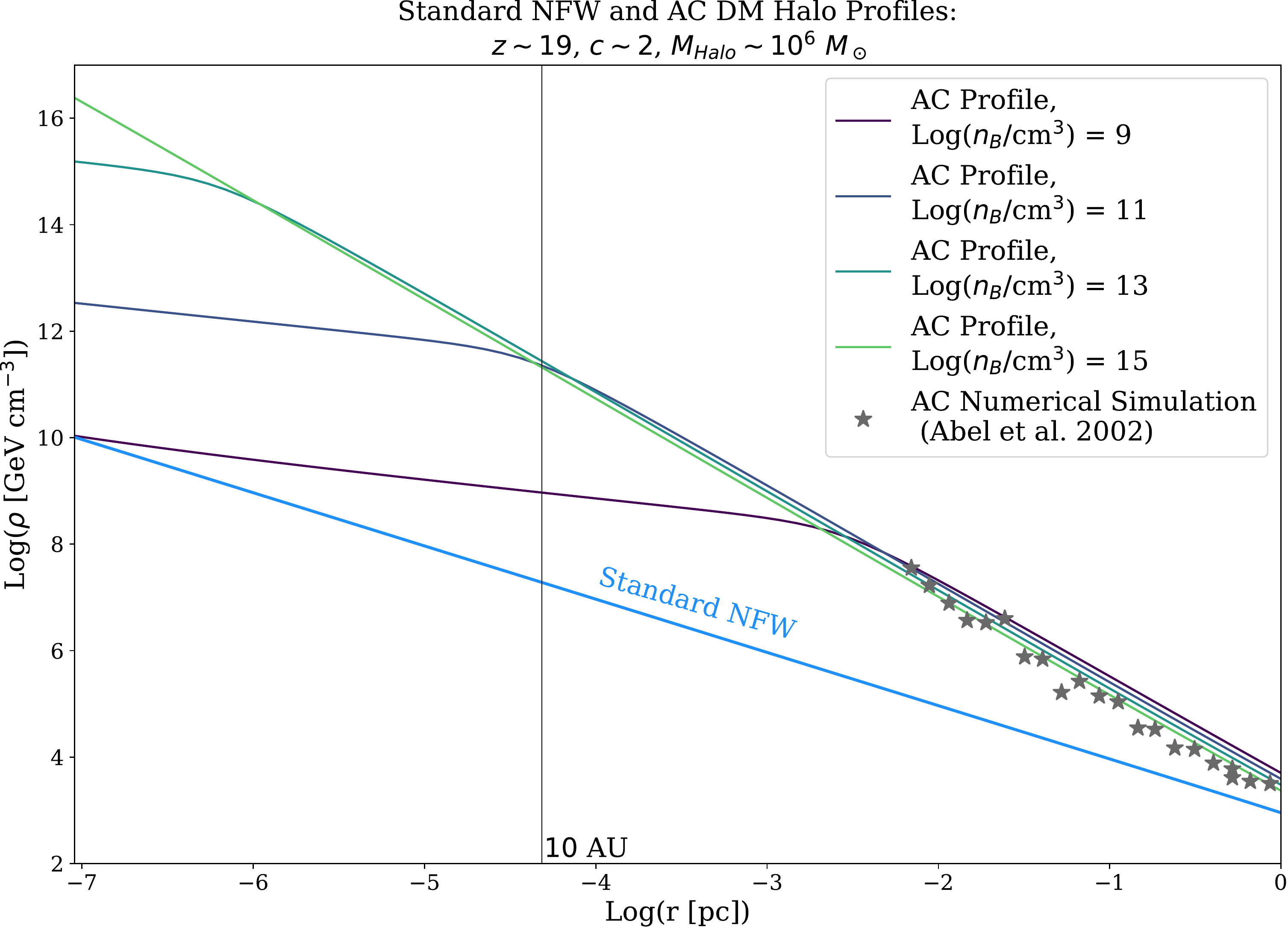}
    \caption{Adiabatically contracted NFW profiles vs. numerical simulation of DM densities during the runaway collapse of a pre-Pop III star molecular gas cloud. Each profile corresponds to a different value for the protostellar core density ($n_B$), labeled in the legend. The simulation data points are taken from Fig.2 of~\cite{Abel:2001}, which corresponds to $n_B\sim 10^{13}\unit{cm}^{-3}$. Resolution limits the simulation from probing the DM densities to scales smaller than $\sim 10^{-2}$~pc. Note the excellent agreement with the AC contracted profile for the same $n_B=10^{13}\unit{cm}^{-3}$.}
    \label{fig:DM_AC}
\end{figure}

 In Fig.~\ref{fig:DM_AC} we show the adiabatically contracted DM densities, obtained using the Blumenthal method~\cite{Blumenthal:1985}, which assumes circular orbits and conservation of angular momentum. Note the response of the DM density profile to the increase in the baryonic core density, $n_B$. The break in the slope of the DM density occurs at the edge of the core ($r_B$). Outside of it, $n_B\sim (r/r_B)^{-2.3}$, whereas inside the core $n_B$ is roughly constant.  Following~\cite{Freese:2008dmdens}, we compare our AC DM densities with those obtained via the numerical simulation of~\cite{Abel:2001}. Numerical resolution limits the simulation from probing DM densities in the inner milliparsec. However, note the excellent agreement of the AC enhanced DM densities vs. the numerical simulation, at a given $n_B$ (in this case $n_B=10^{13}\unit{cm}^{-3}$). An important open question is: up to what $n_B$ does the AC mechanism operate and what will be the feedback effects to stop it? Based on the agreement presented above, we can say that AC operates up to densities of at least $n_B\sim 10^{13}\percc$, and potentially higher.   
 
 Hydrodynamical simulations~\cite{Barkana:2000,Abel:2001,Bromm:2003,Yoshida:2006,Yoshida:2008,Loeb:2010,Grief:2012,Bromm:2013,Klessen:2018} agree that a protostellar core in hydrostatic equilibrium forms when $n_B\sim 10^{22}\unit{cm}^{-3}$, which eventually leads to  one or just a few very massive Pop~III stars, within the inner 10 AU of the DM mini-halo, with the most massive ones closest to the center. As such, a conservative estimate for the ambient DM density relevant for our paper can be approximated with the AC contracted density at roughly $10$ AU, i.e. $5\times 10^{-5}$pc, which, based on Fig.~\ref{fig:DM_AC}, is roughly $\rho_X\sim 10^{13}\GeV\percc$. We will consider $\rho_X$ to range from this conservative value, up to $\rho_X\sim 10^{16}\GeV\percc$, corresponding to a star closer to the center or (and) AC operating to $n_B\gtrsim 10^{13}\percc$.

Going beyond the circular orbit approximation of the Blumenthal method can be done~\cite{Young:1980,Gendin:2004}. Remarkably, the improvements are only up to factors of order unity~\cite{Freese:2008dmdens}. We note these estimates are robust against changes in the initial DM density profile~\cite{Freese:2008dmdens}. In~\cite{Ilie:2020PopIIIa} we show that, up to factors of order unity, the AC estimates for $\rho_X$ are the same for a wide range of concentration parameters for the initial NFW profile ($c\sim 1-10$) and redshifts ($z\sim 5-20$).

In older galaxies, Active Galactic Nuclei, or radiative feedback from very massive stars, can lead to a suppression of the infall of baryons, and therefore a suppression of the enhancement of the DM densities~\cite{Duffy:2010}. Even so,~\cite{Cautun:2020} demonstrates that Milky-Way rotation curve data tends to prefer the physically motivated contracted NFW halo, which  
can be seen as direct experimental evidence of the compression of dark matter densities due to baryonic infall.  

DM annihilation can also have effects on the ambient DM densities, as it can remove DM particles from the environment surrounding a star. Since annihilation is the dominant process at play here (capture and evaporation only happen inside the star), the evolution of $N_X$ in the medium surrounding the star is controlled by the annihilation rate in the vicinity of the star: $\dot{N}_X=-\Gamma_A$. For the three processes considered we have: $\Gamma_A=\int dV n_X^2 \avg{\sigma v}$ (p/s-wave), $\Gamma_A=\int dV n_X^2n_{SM} \avg{\sigma v^2}$ (Co-SIMP DM). As such, one can recast the equation controlling $N_X$ outside of the star into an equation for $\rho_X=n_X m_X$. Since $n_{SM}$ is negligible outside the dense environment of the star, the ambient DM density is not affected by Co-SIMP DM annihilations. However, for p/s-wave DM models we find: $\rho^{-1}_X(t)=\rho^{-1}_{X0}+\rho^{-1}_{AP}(t)$, with $\rho^{-1}_{X0}$ being the initial DM density, and the annihilation plateau (value reached at later times) given by: $\rho_{AP}(t)=m_X/(\avg{\sigma v} t)$.

\begin{figure} [!htb]
    \centering
    \includegraphics[width=0.99\linewidth]{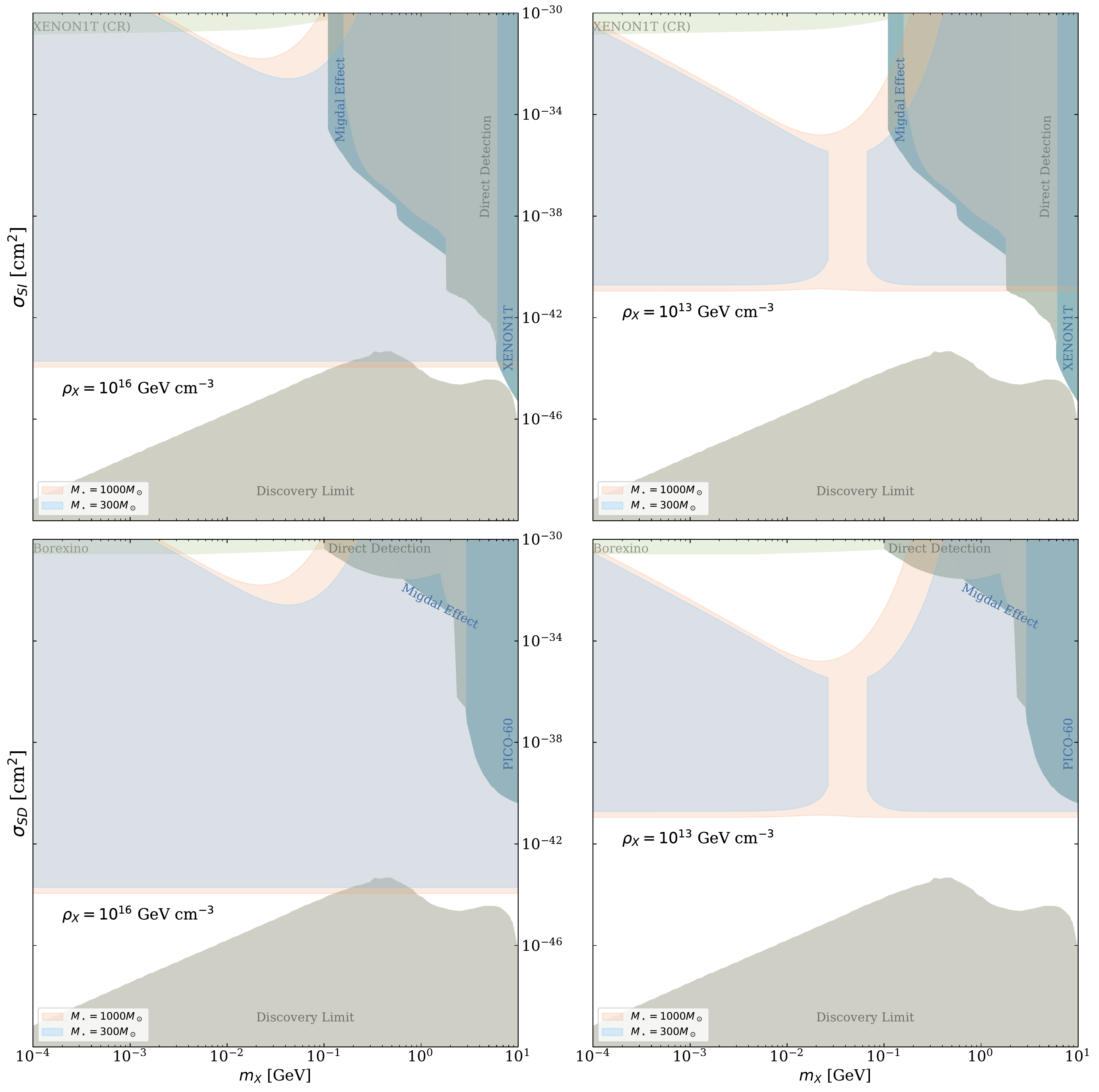}
    \caption{Low $m_X$ DM-proton cross section limits using the $300\Msun$ or the $1000\Msun$ Pop~III stars in the $z\sim 7$ Ly$\alpha$ system found by~\cite{Vanzella:2020}. Top/bottom panels contrast our bounds to most recent exclusion limits for SI~\cite{Aprile:2018,Bringmann:2018cvk,Aprile:2019ldm,Abdelhameed:2019} and SD~\cite{Bringmann:2018cvk,Amole:2019fdf,Aprile:2019ldm,Aprile:2019} interactions from various experiments, each with the name listed inside the corresponding region. Additionally, we plot the limiting region, inaccessible to direct detection experiments, labeled ``Discovery Limit'', i.e. the neutrino floor~\cite{Billard:2013,Battaglieri:2017}. Right/left panels correspond to the two ends of the $\rho_X$ interval considered: $10^{13}-10^{16}\GeV\percc$.}
     \label{fig:SubGeVBounds}
\end{figure}

{\it{DM-proton cross section bounds.   }} 
In this section, we present our main results: constraints on $\sigma$ for SD/SI DM-proton interactions. We start, in Fig.~\ref{fig:SubGeVBounds}, with low mass DM, in particular the Co-SIMP model of~\cite{Smirnov:2020}. We place constraints using either a $300~\Msun$ or a $1000~\Msun$ Pop III star, as identified by~\cite{Vanzella:2020} using  Ly$\alpha$ emission from a $z\sim 7$ system observed in the MUSE deep field data. Note that our excluded regions are only mildly sensitive to the stellar mass, and we get the same bounds for both SD/SI cases. This is in contrast to direct detection experiments on Earth, for which the SD bounds are typically weaker by  about five orders of magnitude. For the highest $\rho_X$ considered ($10^{16}\GeV\percc$), we can probe below the neutrino floor region, for $0.1~\GeV\lesssim m_X\lesssim 1~\GeV$. More importantly, we rule out a large swath of parameter space ($\sigma\gtrsim 10^{-44}\unit{cm}^{2})$ that is currently not constrained. For the more conservative $\rho_X\sim 10^{13}~\GeV\percc$, we are able to rule out a large part of the parameter space above $\sigma\sim 10^{-41}~\unit{cm}^2$. Note that DM evaporation, for the lowest mass star ($100~\Msun$), and in the case of the lower DM density, leads to a loss of sensitivity in narrow funnel region around $10^{-2}~\GeV$. Additionally, our excluded regions are bounded from above by curves that transition between $\sigma\sim m_X^{-2}$, to $\sigma\sim e^{m_X}$.

\begin{figure} [!htb]
    \centering
    \includegraphics[width=0.99\linewidth]{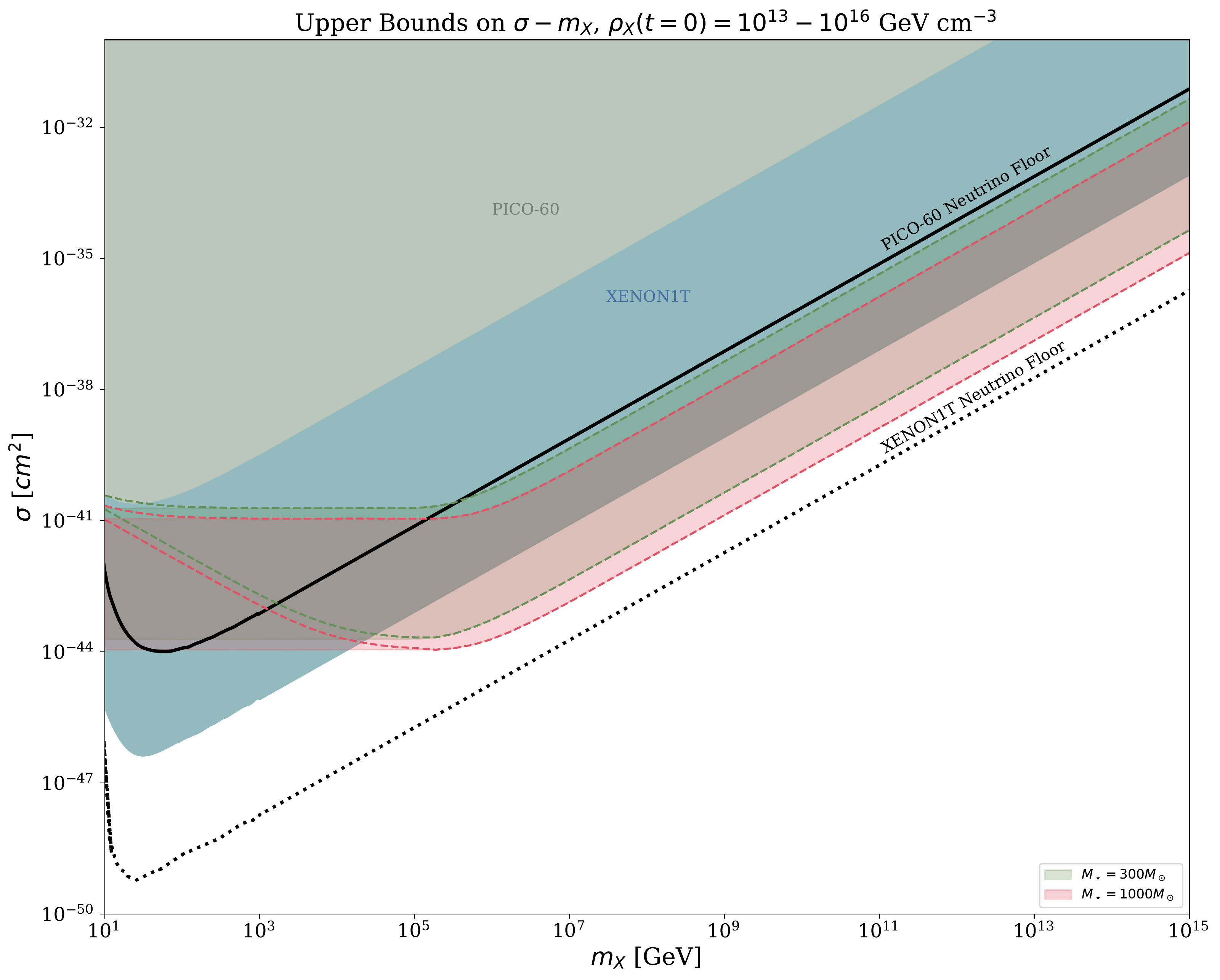}
    \caption{SI/SD DM-proton cross section limits for $m_X\gtrsim 10~\GeV$, obtained using the $300\Msun$ or the $1000\Msun$ Pop~III stars identified by~\cite{Vanzella:2020}. For each star the exclusion limit is represented by a band, corresponding to $10^{13}\GeV\percc\lesssim\rho_X\lesssim10^{16}\GeV\percc$. The dashed colored lines (green/red) correspond to the exclusion limits we get, when including the effects of the annihilation plateau at $t\sim1$~Gyrs for ambient DM densities. The most recent SD bounds from direct detection are taken from the PICO experiment~\cite{Amole:2019fdf}, whereas X1T~\cite{Aprile:2018} is used the SI case. Additionally we plot the the boundaries of the neutrino floor regions for XENON~\cite{Billard:2013,Battaglieri:2017} (relevant for SI bounds) and for the PICO detector~\cite{Ruppin:2014}(relevant for SD bounds).}
     \label{fig:bounds}
\end{figure}

In Fig.~\ref{fig:bounds}, we present our bounds on the DM-proton scattering cross section, contrasted against the current, deepest available exclusion limits from direct detection experiments, for $m_X\gtrsim 10\GeV$. Even for the conservative $\rho_X\sim 10^{13}~\GeV\unit{cm}^{-3}$, all of our exclusion limits rule out a large swath of parameter space for SD DM-proton cross sections that has yet to be explored by direct detection experiments. For the SI case, if $\rho_X\sim10^{14}\GeV\percc$, our bounds are competitive with those obtained by X1T, at $m_X\gtrsim10^{5}\GeV$, whereas for any higher $\rho_X$, we begin to probe regions of parameter space currently unexplored by direct detection experiments. Finally, we point out that for a Pop~III star of any given mass, there is a corresponding DM density ($\rho_{X;\text{nf}}$) for which the mere existence of the star in question will rule out DM-proton cross sections all the way down to the neutrino floor. At the same stellar mass (or $\rho_X$), a higher $\rho_X$ (or $\Mstar$) implies probing below the neutrino floor. For example, whenever $\rho_X\gtrsim10^{18}~\GeV\unit{cm}^{-3}$, the identification of {\it{any}} $100~\Msun$ would probe DM-proton cross sections below the neutrino floor. In the case of a $1000\Msun$ Pop~III star, the corresponding $\rho_{X;\text{nf}}\sim 10^{17}~\unit{GeV}\unit{cm}^{-3}$.

In summary, we demonstrated that the observation of Pop~III stars can be used to place strong constrains on the DM-proton cross section. Applying our method to the candidate Pop~III system at $z\sim 7$~\cite{Vanzella:2020}, we obtain some of the most stringent bounds to-date, at masses outside of the WIMP DM window. For SD interactions, we are able to probe well below the neutrino floor of the most sensitive current direct detection experiment, PICO. Followup observations with JWST are necessary to confirm the Pop~III nature of the system we used here, and therefore the limits we obtained. 

{\it{Aknowledgements.     }} CI would like to thank Katherine Freese and Paolo Gondolo for sharing the code we used to calculate the adiabatically contracted NFW profiles used here. This is the same code used in ~\citet*{Spolyar:2008dark}, where the conditions for the formation Dark Stars were first identified. CL thanks the financial support from Colgate University, via the Research Council student wage grant, and the Justus ’43 and Jayne Schlichting Student Research Funds.

\bibliography{RefsDM}

\end{document}